\documentclass[twocolumn,aps,pra,groupedaddress,nobibnotes,showpacs]{revtex4}
\usepackage{graphicx}

\begin{document}

\title{Single-photon two-qubit ``entangled'' states: preparation and measurement}

\author{Yoon-Ho Kim}\email{kimy@ornl.gov}
\affiliation{Center for Engineering Science Advanced Research, Computer Science \& Mathematics Division, Oak Ridge National Laboratory, Oak Ridge, Tennessee 37831}

\date[]{to appear in Physical Review A, Rapid Communication (2003)}

\begin{abstract}
We implement experimentally a deterministic method to prepare and measure so called single-photon two-qubit entangled states or single-photon Bell-states, in which the polarization and the spatial modes of a single-photon each represent a quantum bit. All four single-photon Bell-states can be easily prepared and measured deterministically using linear optical elements alone. We also discuss how this method can be used for recently proposed single-photon two-qubit quantum cryptography protocol.
\end{abstract}


\maketitle

Entanglement usually refers to multi-particle quantum entanglement which exhibits non-local quantum correlations that are verified experimentally by observing multi-particle quantum interference \cite{epr}. For example, the two-photon entanglement in spontaneous parametric down-conversion (SPDC) photon pair is observed in the form of fourth-order quantum interference \cite{shih}. Recently, multi-particle entanglement and quantum interference effects have been shown to be essential for new quantum applications, such as quantum information, metrology, lithography, etc. \cite{nielson,metrology,litho}.

A different type of ``entanglement'', namely ``single-particle entanglement''  or ``entanglement'' of internal degrees of freedom of a single quantum particle started to attract interest recently. Although single-particle entanglement lacks non-locality which is at the heart of multi-particle entanglement necessary for a number of quantum applications mentioned above \cite{analogy}, it has been shown nevertheless to be useful for simulating certain quantum algorithms at the expense of exponential increase of required physical resources \cite{opticallogic,noent,englert}. Several experiments illustrating this point have already been carried out \cite{waveexp,atomexp}. It has also been shown that single-photon two-qubit states may be useful for deterministic cryptography \cite{beige}. 

For single-photon two-qubit states, two dichotomic variables of a single-photon represent the two qubits. Usually, one is the polarization qubit in which the basis states are the orthogonal polarization states of the single-photon (e.g., horizontal $|H\rangle$ or vertical $|V\rangle$ polarization states) and the other is the spatial qubit in which the basis states are two spatial modes of the single-photon (e.g., the photon travels in path $a$ or in path $b$) \cite{analogy,opticallogic,englert,beige}. Clearly, a complete basis for the single-photon two-qubit state can be formed by a set of any four orthonormal states of the photon. For example, a set of $|a,V\rangle$, $|a,H\rangle$, $|b,V\rangle$, and $|b,H\rangle$ forms a complete (product) basis for the single-photon two-qubit Hilbert space. Preparation and measurement of such (product) basis states are trivial since interference is not required in both preparation and measurement stages. It is also possible to consider the product basis states which are composed of symmetric and antisymmetric superposition states of the polarization qubit and the spatial qubit \cite{beige}. We will discuss this case later in this paper.

In the entangled basis of the single-photon two-qubit state, the single-photon Bell-states
\begin{eqnarray}
|\Psi^{(\pm)}\rangle &\equiv& \frac{1}{\sqrt{2}} ( |a,H\rangle \pm |b,V\rangle ),\nonumber\\
|\Phi^{(\pm)}\rangle &\equiv& \frac{1}{\sqrt{2}} ( |a,V\rangle \pm |b,H\rangle ),\nonumber
\end{eqnarray}
form a complete basis. In this paper, we propose a deterministic method to prepare and measure the ``single-photon Bell-states,'' report results on the experimental implementation of the method, and discuss some potential difficulties related to the single-photon two-qubit quantum cryptography proposed in Ref.~\cite{beige}. 

\begin{figure}[t]
\includegraphics[width=3.3in]{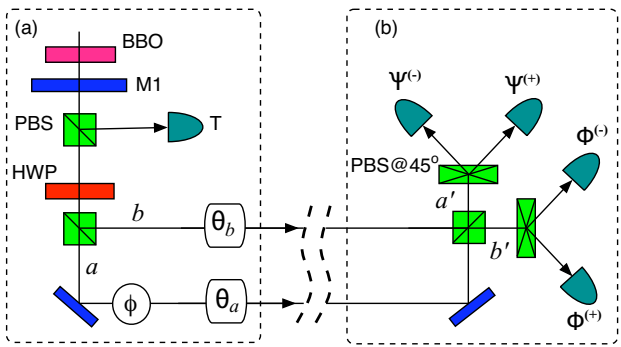}
\caption{\label{fig:setup}Outline of the experiment. (a) Preparation of a single-photon two-qubit entangled state (Bell-state) is done by using a single-photon polarized in $45^\circ$ and a PBS. The single-photon state is prepared by detecting one photon of the SPDC photon pair with the trigger detector T. HWP rotates the polarization of the single-photon to $45^\circ$ and the second PBS prepares a single-photon Bell-state. Additional phase and polarization elements, $\phi$, $\theta_a$, and $\theta_b$, may be used to prepare the other three Bell-states. (b) Bell-basis measurement. A PBS, which mixes spatial modes $a$ and $b$, is followed by a $45^\circ$ oriented PBS located at each output port. The detector click at the outputs of $45^\circ$ oriented PBS' uniquely identify four single-photon Bell-states. The preparation and measurement stages together form an equal-path Mach-Zehnder interferometer.}
\end{figure}

The outline of the experimental setup is shown in Fig.~\ref{fig:setup}. Let us first focus on the state preparation part shown in Fig.~\ref{fig:setup}(a). The single-photon state used in this experiment was generated using the post-selection method first demonstrated in Ref.~\cite{hong}. A 2 mm thick type-II BBO crystal was pumped with a 351.1 nm argon laser. Orthogonally polarized spontaneous parametric down-conversion (SPDC) photon pairs generated in the crystal had central wavelength of 702.2 nm and propagated collinearly with the pump beam. After removing the pump laser beam with a dichroic mirror M1, the vertically polarized photon was directed to the trigger detector T by a polarizing beamsplitter PBS and the trigger signal indicated that there was one and only one photon (polarized horizontally) traveling in the other output ports of PBS \cite{note1}. Note that this kind of post-selected single-photon states have recently been used for demonstrating linear optical quantum logic gates and memory \cite{pittman}. 

A half-wave plate HWP orieinted at $22.5^\circ$ rotated the polarization of the horizontally polarized single-photon to $45^\circ$ polarization state just before the second PBS. After the second PBS, the state of the single-photon can be written as 
$$
|\Psi^{(+)}\rangle=(|a,H\rangle+|b,V\rangle)/\sqrt{2},
$$ 
which is a single-photon Bell-state. The other three single-photon Bell-states can be prepared by using an additional phase shifter $\phi$ and polarization rotating half-wave plates $\theta_a$ and $\theta_b$. The spatial phase $\phi$ can be introduced, for example, by slightly moving the mirror in path $a$ and it determines whether the two amplitudes in a single-photon Bell-state interferes constructively ($\phi=0$) or destructively ($\phi=\pi$).
 The half-wave plates, indicated as $\theta_a$ and $\theta_b$, inserted in the beam paths $a$ and $b$ may either flip the polarization ($\theta_a=\theta_b=45^\circ$) or do nothing ($\theta_a=\theta_b=0^\circ$). Therefore, all four single-photon Bell-states can be easily prepared in this setup by suitable combinations of the spatial phases and polarization flip. 

Note that it is also possible to prepare single-photon two-qubit product states using this scheme. To prepare single-photon two-qubit product states in the basic qubit bases ($|a\rangle$, $|b\rangle$, $|H\rangle$, and $|V\rangle$), we need to control the HWP orientations ($0^\circ$ or $45^\circ$, but not $22.5^\circ$) and the angles of $\theta_a$ and $\theta_b$ ($0^\circ$ or $45^\circ$). For example, $0^\circ$ HWP angle and $\theta_a=0^\circ$ prepares the state $|a,H\rangle$. The spatial phase $\phi$ is not relevant in this case because interference does not play any role. 

We can also prepare single-photon two-qubit product states in a superposition basis, as proposed in the single-photon two-qubit quantum cryptography protocol \cite{beige},
$$
(|a,+45^\circ\rangle,|b,-45^\circ\rangle,|S,V\rangle, |A,H\rangle),
$$
where
$$
	\left.
	\begin{array}{l}
		|S\rangle\\
		|A\rangle
	\end{array}
	\right\} = \frac{1}{\sqrt{2}}(|a\rangle \pm |b\rangle), \hspace{.3cm} 
	\left.
	\begin{array}{l}
		|+45^\circ\rangle\\
		|-45^\circ\rangle
	\end{array}
	\right\} = \frac{1}{\sqrt{2}}(|H\rangle \pm |V\rangle).
$$
For example, $|S,V\rangle$ state can be prepared by setting the HWP at $22.5^\circ$, $\theta_a=45^\circ$, and $\theta_b=0^\circ$. 

This setup therefore allows preparation of various single-photon two-qubit states, as required for the deterministic quantum cryptography proposed in Ref.~\cite{beige}. Since all the phase and polarization adjusting components in this setup can be replaced with electro-optical devices, automated random switching among different states is possible.

Let us now discuss the measurement of single-photon Bell-states. A complete measurement of two-photon polarization Bell-states requires both nonlinear optical effects and quantum interference \cite{bsm,tele}. On the other hand, a complete measurement of the single-photon Bell-states require only single-photon interference effects and linear optical elements \cite{boschi}. It is because entangling or interacting two separate photons requires nonlinear optical elements, but ``entangling-unentangling'' single-photon two-qubit states require only linear optical elements as we have seen earlier.

The single-photon two-qubit Bell-basis measurement scheme is shown in Fig.~\ref{fig:setup}(b). First, we mix the spatial qubit modes, labeled as $a$ and $b$, at a polarizing beamsplitter. Since the single-photon Bell-basis detector relies on the single-photon interference effect, it is necessary that the paths $a$ and $b$ are kept equal. The polarizing beamsplitter transforms the incident amplitudes in the following way:
\begin{eqnarray}
|a,H\rangle &\rightarrow& |a',H\rangle, \hspace{1cm} |b,V\rangle \rightarrow |a',V\rangle, \nonumber\\
|b,H\rangle &\rightarrow& |b',H\rangle, \hspace{1cm} |a,V\rangle \rightarrow |b',V\rangle.\nonumber
\end{eqnarray}
The single-photon Bell-states are then transformed by the polarization beamsplitter,
\begin{eqnarray}
|\Psi^{(\pm)}\rangle &\rightarrow& \frac{1}{\sqrt{2}}|a'\rangle ( |H\rangle \pm |V\rangle ) = \left\{ \begin{array}{l}
                     |a'\rangle|+45^\circ\rangle\\
                     	|a'\rangle|-45^\circ\rangle
					\end{array}
				\right.,\nonumber\\
|\Phi^{(\pm)}\rangle &\rightarrow& \frac{1}{\sqrt{2}}|b'\rangle ( |V\rangle \pm |H\rangle ) =\left\{ \begin{array}{l}
                     |b'\rangle|+45^\circ\rangle\\
                     	-|b'\rangle|-45^\circ\rangle
					\end{array}
				\right..\nonumber
\end{eqnarray}
Clearly, a $45^\circ$ oriented polarizing beamsplitter (PBS@$45^\circ$) inserted at modes $a'$ and $b'$ can separate the above states into four distinct spatial modes. The four single-photon detectors placed at the output ports of PBS@$45^\circ$, shown in Fig.~\ref{fig:setup}(b), therefore produce an unambiguous signal which corresponds to the input single-photon Bell-state.

\begin{figure}[t]
\includegraphics[width=3.3in]{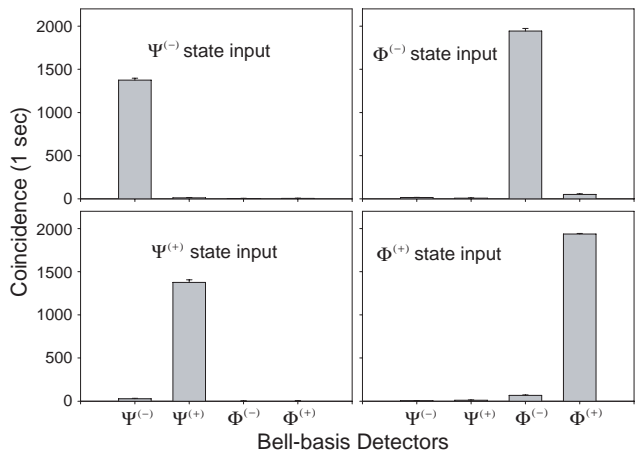}
\caption{\label{fig:data}Experimental data showing the outputs of Bell-state detectors for a given Bell-state input. Error counts show up due to the result of imperfect experimental alignment and phase instabilities. $|\Psi^{(\pm)}\rangle$ detectors show lower count rate than $|\Phi^{(\pm)}\rangle$ detectors due to lower photon coupling efficiencies. }
\end{figure}

We have implemented experimentally the preparation and measurement scheme for the single-photon Bell-states. The argon pump laser was approximately 200 mW. The distance from the second polarizing beamsplitter (which is used to prepare the single-photon Bell-state) to the third polarizing beamsplitter (which is used to measure the state) was about 50 cm. $\theta_a$ and $\theta_b$ angles were set by hand and $\phi$ was set by moving the mirror in the beam path $a$ with a computer controlled DC motor. In the Bell-basis measurement part of the setup, a half-wave plate and a polarizing beamsplitter were used instead of rotating the PBS by $45^\circ$. A multi-mode fiber coupled single-photon counting detector was placed at each output modes and the coincidence between the trigger detector T and the Bell-basis detector was measured. The coincidence window in this experiment was about 3 nsec. 

Figure~\ref{fig:data} shows the output of the Bell-basis detectors for four different single-photon Bell-states as the input and it clearly indicates that the Bell-basis detectors behave as expected: e.g., for $|\Psi^{(-)}\rangle$ state input, only $|\Psi^{(-)}\rangle$ detector produces a signal. The data however show that there are still some probabilities that the single-photon ends up at a wrong Bell-basis detector. These errors are the result of imperfect alignment of the experimental setup and phase instabilities of the overall Mach-Zehnder interferometer. Also, $|\Psi^{(\pm)}\rangle$ detectors show lower count rate than $|\Phi^{(\pm)}\rangle$ detectors. This is due to different photon coupling efficiencies of fiber-coupled single-photon detectors. 

In this experiment, the coincidence counts between the trigger detector T and four Bell-basis detectors are measured so the dark counts of individual detectors did not show up in the data. Similar reduction of dark counts can be expected in real-world situations as well, if the single-photon source is pulsed and the Bell-basis detectors are gated accordingly. 

\begin{figure}[t]
\includegraphics[width=3.in]{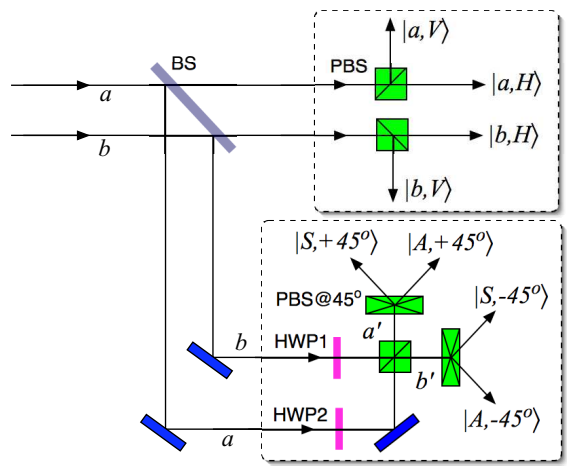}
\caption{\label{fig:proposal}Possible receiver design for proposed single-photon two-qubit quantum cryptography protocol. Half-wave plates (HWP1 oriented at $22.5^\circ$ and HWP2 oriented at $67.5^\circ$) transforms the $(|B'_i\rangle)$ states to the Bell-states with one to one correspondence. }
\end{figure}

Let us now briefly discuss how the single-photon Bell-basis detection scheme demonstrated in this paper might be applied for the single-photon two-qubit quantum cryptography  proposed in Ref.~\cite{beige}. In their scheme, the receiver must have two detection bases, for example, 
\begin{eqnarray}
(|B_i\rangle)&=&(|a,V\rangle, |a,H\rangle, |b,V\rangle, |b,H\rangle),\nonumber\\
(|B'_i\rangle)&=&(|S,+45^\circ\rangle, |A,+45^\circ\rangle, |S,-45^\circ\rangle, |A,-45^\circ\rangle).\nonumber
\end{eqnarray}
The $(|B_i\rangle)$ basis detection is trivial since no interference is required for the state detection: polarizing beamsplitters in paths $a$ and $b$ would do the job. For $(|B'_i\rangle)$ basis detection, however, interference is required. We have seen earlier that the Bell-basis detection scheme produces a unique and unambiguous output signal corresponding to the input single-photon Bell-state. Therefore, for unambiguous and complete $(|B'_i\rangle)$ basis detection, we only need to find a way to transform the $(|B'_i\rangle)$ basis states to the Bell-states with one-to-one correspondence in one optical setting. This may be accomplished by introducing a half-wave plate at each input port of the the polarizing beamsplitter in the Bell-basis detection scheme. If the half-wave plate in path $a$ input is oriented at $67.5^\circ$ and the half-wave plate in path $b$ is oriented at $22.5^\circ$, the $(|B'_i\rangle)$ states are transformed to the Bell-states with exact one to one correspondence just before the polarizing beamsplitter,
\begin{eqnarray}
&|S,+45^\circ\rangle& \rightarrow |\Psi^{(+)}\rangle, \hspace{0.5cm} |A,+45^\circ\rangle \rightarrow |\Psi^{(-)}\rangle,\nonumber\\ 
&|S,-45^\circ\rangle& \rightarrow |\Phi^{(+)}\rangle, \hspace{0.5cm} |A,-45^\circ\rangle \rightarrow |\Phi^{(-)}\rangle.\nonumber
\end{eqnarray}
Complete and unambiguous $(|B'_i\rangle)$ basis detection is therefore possible with a very simple modification of the Bell-basis detection scheme. A possible receiver design for proposed single-photon two-qubit quantum cryptography can be seen in Fig.~\ref{fig:proposal}.

Finally, we discuss some potential difficulties regarding quantum cryptography schemes using the single-photon two-qubit state. As we have seen in this paper, the sender and the receiver in quantum cryptography schemes using the single-photon two-qubit state are two sub-divisions of a huge Mach-Zehnder interferometer. Since single-photon interference is critical for reliable and unambiguous Bell-state detections, the two spatial qubit modes $a$ and $b$ should be free of any phase fluctuations. If, for example, a $\pi$ relative phase is temporarily introduced between the two spatial modes when the photons are traveling between the two parties, the Bell-state detectors will produce an incorrect output: for example, $|\Psi^{(-)}\rangle$ state sent to the receiver will trigger $|\Psi^{(+)}\rangle$ detector instead. Phase stability, in addition to polarization stability, therefore will be a serious issue for implementing long distance quantum communication using the single-photon two-qubit state. 

In summary, we designed and implemented a deterministic method to prepare and measure all four ``single-photon two-qubit entangled states'' or ``single-photon Bell-states.'' Although the quantum cryptography protocol based on single-photon two-qubit states promises to give a factor of 2 gain compared to other cryptography scheme \cite{beige}, such as the one based only on single-photon polarization states \cite{bb}, the price to pay is substantially higher maintenance cost to ensure that the spatial modes are free of any phase instabilities, in addition to the polarization stability.  

The author acknowledges many useful discussions with C. D'Helon, W.P. Grice, and V. Protopopescu. This research was supported in part by the U.S. DOE, Office of Basic Energy Sciences and the National Security Agency. The Oak Ridge National Laboratory is managed for the U.S. DOE by UT-Battelle, LLC, under contract No.~DE-AC05-00OR22725.


\begin{thebibliography}{}

\bibitem{epr} A. Einstein, B. Podolsky, and N. Rosen, Phys. Rev.
\textbf{47}, 777 (1935); J.S. Bell, \textit{Speakable and Unspeakable in Quantum
Mechanics} (Cambridge University Press, New York, 1987).

\bibitem{shih} Y.H. Shih and C.O. Alley, in \textit{Proceedings of the Second International Symposium on Foundations of Quantum Mechanics in the Light of New Technology, Tokyo, 1986}, edited by M. Namiki \textit{et al}. (Physical Society of Japan, Tokyo, 1987); Phys. Rev. Lett. \textbf{61}, 2921 (1988); Z. Y. Ou and L. Mandel, \textit{ibid}. \textbf{61}, 50 (1988).

\bibitem{nielson} M.A. Nielsen and I.L. Chuang, \textit{Quantum Computation and Quantum Information}, (Cambridge Univ. Press, 2000).

\bibitem{metrology} D.N. Klyshko, Sov. J. Quant. Electron. \textbf{7}, 591 (1977); A. Migdall \textit{et al.}, Appl. Opt. \textbf{37}, 3455 (1998); J. Dowling, Phys. Rev. A \textbf{57}, 4736 (1998); V. Giovannetti, S. Lloyd, and L. Maccone, Phys. Rev. A \textbf{65}, 022309 (2002).

\bibitem{litho} A.N. Boto \textit{et al.}, Phys. Rev. Lett. \textbf{85}, 2733 (2000); M. D'Angelo, M.V. Chekhova, and Y. Shih, Phys. Rev. Lett. \textbf{87}, 013602 (2001).

\bibitem{analogy} R.J.C. Spreeuw, Found. Phys. \textbf{28}, 361 (1998); Phys. Rev.
A \textbf{63}, 062302 (2001).

\bibitem{opticallogic} N.J. Cerf, C. Adami, and P.G. Kwiat, Phys. Rev. A
\textbf{57}, R1477 (1998); G. Brassard, S.L. Braunstein, and R. Cleve, Physica
D \textbf{120}, 43 (1998).

\bibitem{noent} S. Lloyd, Phys. Rev. A \textbf{61}, 010301(R) (1999);
P. Knight, Science \textbf{287}, 441 (2000).

\bibitem{englert} B.-G. Englert, C. Kurtsiefer, and H. Weinfurter, Phys. Rev. A \textbf{63}, 032303 (2001).

\bibitem{waveexp} P.G. Kwiat \textit{et al.}, J. Mod. Opt. \textbf{47}, 257
(2000); C. Dorrer \textit{et al.}, QELS 2001 Proceedings, QWB3 (2001); N.
Bhattacharya \textit{et al.}, Phys. Rev. Lett \textbf{88}, 137901 (2002).

\bibitem{atomexp} J. Ahn, T.C. Weinacht, and P.H. Bucksbaum, Science \textbf{287},
463 (2000); D.A. Meyer, Science \textbf{289}, 1431a (2000); P.G. Kwiat and R.J.
Hughes, \textit{ibid}.

\bibitem{beige} A. Beige, B.-G. Englert, C. Kurtsiefer, and H. Weinfurter, J. Phys. A: Math. Gen. \textbf{35}, L407 (2002).

\bibitem{hong} C.K. Hong and L. Mandel, Phys. Rev. Lett. \textbf{56}, 58 (1986).

\bibitem{note1} One can confirm post-selected single-photon states in SPDC experimentally by directly measuring the post-selected photon statistics \cite{hong} or by measuring the anticorrelation parameter \cite{grangier}.    

\bibitem{grangier} P. Grangier, G. Roger, and A. Aspect, Europhys. Lett. \textbf{1}, 173 (1986).

\bibitem{pittman} T.B. Pittman, B.C. Jacobs, and J.D. Franson, Phys. Rev. A \textbf{66}, 042303 (2002); T.B. Pittman and J.D. Franson, Phys. Rev. A \textbf{66}, 062302 (2002).

\bibitem{bsm} L. Vaidman and N. Yoran, Phys. Rev. A \textbf{59}, 116 (1999); N. Lutkenhaus, J. Calsamiglia, and K.-A. Suominen, \textit{ibid.}, \textbf{59}, 3295 (1999).

\bibitem{tele} Y.-H. Kim, S.P. Kulik, and Y. Shih, Phys. Rev. Lett.
\textbf{86}, 1370 (2001); J. Mod. Opt. \textbf{49}, 221 (2002).

\bibitem{boschi} This effect has been used in a recent demonstration of quantum teleportation by D. Boschi \textit{et al.} [Phys. Rev. Lett. \textbf{80}, 1121 (1998)]. The qubit to be teleported, however, cannot come from outside in this case.

\bibitem{bb} C.H. Bennett and G. Brassard, \textit{Proc. IEEE Int. Conf. on Computers, Systems, and Signal Processing} (IEEE, New York, 1984); C.H. Bennett, Phys. Rev. Lett. \textbf{68}, 3121 (1992).

\end{thebibliography}
\end{document}